\documentclass[12pt]{article}

        \usepackage{amsfonts}
        \usepackage{amssymb}

\def\be{\begin{eqnarray}}
\def\ee{\end{eqnarray}}
\def\bee{\begin{eqnarray*}}
\def\eee{\end{eqnarray*}}

\def\raw{\rightarrow}

\def\ep{\epsilon}

         \def\tr{\hbox{tr}}

\def\rt2{ \frac{1}{\sqrt{2}} }

\def\raw{\rightarrow}

\def\dim{{\rm dim }}
\def\conv{{\rm conv }}
\def\vol{{\rm vol }}

\def\cs{{\mathcal{S}}}
\def\cd{{\mathcal{D}}}
\def\calm{{\mathcal{M}}}
\def\calb{{\mathcal{B}}}
\def\calh{{\mathcal{H}}}
\def\cale{{\mathcal{E}}}

\def\wrt{with respect to }

\title{The volume of separable states is super-doubly-exponentially small}

\author {
  Stanislaw Szarek (Paris and Cleveland) }

\date{}

\begin{document}

  \maketitle

\begin{abstract}
In this note we give sharp estimates on the volume of the set of separable
states on $N$ qubits. In particular, the magnitude of the ``effective radius"
of that set in the sense of volume is determined up to a factor which is a 
(small) power of $N$, and thus precisely on the scale of powers of its
dimension.  
Additionally, one of the appendices contains new sharp estimates 
(by known methods) for the expected values of norms of the 
GUE random matrices.  We employ standard tools of classical convexity, 
high-dimensional probability and geometry of Banach spaces.
\end{abstract}

\vskip.5cm
Let $\calh = \calh_N := ({\mathbb {C}}^2)^{\otimes N}$ be the $N$-fold tensor
power  of ${\mathbb {C}}^2$ and denote by $d =2^N$ its dimension. 
In this note we investigate the structure of 
the set $\mathcal{D} =\mathcal{D}_N= \mathcal{D}(\calh_N)$ of states on 
$\calb(\calh_N)$   
and, in particular, of its subset $\mathcal{S} = \mathcal{S}_N$ consisting 
of (mixtures of) {\em separable } states. 
 We recall that when
$\mathcal{D}$ is identified with the set of density matrices
$\{\rho \in \calb(\calh) : \rho \mbox{ is positive semi-definite and } 
\tr \rho = 1\}$, then 
$$
\mathcal{S} = \conv \{\rho_1 \otimes \rho_2 \otimes \ldots \otimes \rho_N  :
\rho_j \in \mathcal{D}({\mathbb {C}}^2), j=1,2, \ldots, N \} .
$$ 
Above and in what follows, we skip the subscript $N$ whenever its value
is clear from the context.  We emphasize that separability of a state 
on $\calb(\calh)$ is not an intrinsic property of the Hilbert space $\calh$ 
or the algebra $\calb(\calh)$; it
{\em does } depend on the particular decomposition of $\calh$ as a tensor
product of (smaller) Hilbert spaces.

The question of the size of $\mathcal{S}$ and, particularly, of its relative
size as a subset of $\mathcal{D}$ was raised in \cite{ZHSL} and further 
investigated, among others, in \cite{BCJLPS, VT, pr, qudits, GB} (see
also the survey \cite{prLAA}). One of the parameters that have been studied was
the maximal  size of homothetic images of
$\mathcal{D}$ contained in
$\mathcal{S}$.  More precisely, one asks for which values of $\ep$ 
(say, with $\ep > 0$) we have
\begin{equation} \label{homo}
\ep \mathcal{D} + (1-\ep ) I_d/d \subset \mathcal{S} , 
\end{equation} 
where $I_d$ stands here for the identity matrix 
in $d$ dimensions; in the present context, $I_d/d$ is referred to as 
``the maximally mixed state." 
Alternatively, one considers inclusions of type (\ref{homo}) with
$\mathcal{D}$  replaced by the appropriate Euclidean (Hilbert-Schmidt) ball
$B$.  The bounds obtained to date show that, in both cases, the optimal
(largest)  value of $\ep$ is (asymptotically, as $d \rightarrow \infty$) of
order contained between
$d^{-1}$ and
$d^{-3/2}$.  While clarifying the situation somewhat,
all these results  leave open the question  of the precise asymptotic order of
the ``in-radii" of
$\mathcal{S}$ on the power  scale in $\dim \, \mathcal{S} = \dim \, \mathcal{D}
= d^2-1$ as well as the issue of  the ``size" of $\mathcal{S}$ when measured by
global invariants such as volume.  In the latter direction we obtain
here the following bounds
\begin{equation} \label{vr}
\frac{c}{d^{1/2 +\alpha}} \  \le \  
\left( \frac{{\rm vol \,} \mathcal{S}}{{\rm vol \,} \mathcal{D}} 
\right)^{1/{\dim \, \mathcal{S}}} 
\le \  \frac{C (\log{d} \log{\log{d}})^{1/2}}{d^{1/2+\alpha}} \  ,
\end{equation}
where $c, C>0$ are universal effectively computable numerical constants 
and $\alpha :=\log_2{(27/16)}/8 \approx 0.094361$ (in other words,
$d^{1/2+\alpha} = 3^{3N/8}$). 
The ``effective radius" of $\mathcal{S}$  in the sense of volume is thus 
precisely determined on the scale of powers of $d$. Since complexity of 
a set can often be estimated using volumetric methods (see \cite{pisier} for 
a modern exposition of this circle of ideas), this goes a long way  
towards the ability to compare complexities of $\mathcal{S}$ and of 
$\mathcal{D}$ -- even though the so-called Bures metric and the related 
volume may be more appropriate measures of size in the present context
(see \cite{BC}).
In what follows we shall present the main line of the argument leading to 
(\ref{vr}), relegating to Appendices the discussion of some 
peripheral issues as well as the description of results and  
concepts from convexity and geometry of Banach spaces that are being used.
We refer to \cite{BCJLPS, HHH, JL, GB}
for a more professional exposition of the relevance of separability to 
quantum computation in general and to NMR computing in particular. 

For comparison, we note that the results of \cite{BCJLPS}  
and \cite{GB} implied {\em lower} estimates on 
$({\rm vol \,} \mathcal{S}/{\rm vol \,}
\mathcal{D})^{1/{\dim \, \mathcal{S}}}$ 
which were of order $d^{-\beta}$ with, 
respectively, $\beta = \log{10}/\log{4} \approx 1.660964$ and $\beta = 1$.
By contrast,  no  non-trivial {\em upper} estimates on the volume of
${\mathcal{S}}$ were apparently available, except in very low dimensions.   
In the opposite direction, the expression on the right hand side of  (\ref{vr})
yields upper estimates on the $\ep$'s that may work in (\ref{homo})
and related inclusions. However,  (\ref{vr}) does not generally improve 
on the most recent prior results of that type. One case when it does is by
giving  an upper bound on the radius of a Euclidean ball that may be contained
in ${\mathcal{S}}$ which is tighter (roughly, by a factor of $d^\alpha$) than
the rather elementary $O(d^{-1})$ estimate; see (\ref{vrSigma}) and Appendix H
for more explicit statements in this regard and for more comments. Here we will
just mention that our method does not give -- at least without any  additional
work -- any explicit state that constitutes an obstruction  to the inclusion
$\ep B + (1-\ep ) I_d/d \subset \mathcal{S}$ for  $\ep = o(d^{-1})$, and that
our results suggest that it may be more appropriate to compare
${\mathcal{S}}$  to an ellipsoid which is substantially different from the one
induced by the Hilbert-Schmidt norm  (see the paragraph containing
(\ref{killing})).

Since it is conceivable that the inequalities (\ref{vr}) may be of interest not
just asymptotically, but also for some specific ``moderately large" 
values of $N$, we put some effort into obtaining reasonable (but certainly not
optimal) values of the numerical constants. Our main argument gives 
$c=1/4$ and shows that (\ref{vr}) holds with 
$4(N\log_2{(4N)})^{1/2}= 4(\log_2{d} \, \log_2{(4 \log_2{d})})^{1/2}$
in the numerator of its third member.  A slightly more precise (and more
tedious) calculation yields
$c=\sqrt{e/8\pi}
\approx 0.32887$; see the comments following (\ref{vrSigmamod}) and Appendix E. 
It is also easy to follow the argument and to obtain  somewhat
sharper estimates for specific values of
$N$, which may be of interest, e.g., in the context of a threshold of 
23 mentioned in \cite{GB}. Such improvements are sketched in Appendix G
leading to a non-trivial (i.e., $<1$) bound on 
$({\rm vol \,} \mathcal{S}/{\rm vol \,} \mathcal{D})^{1/{\dim \, \mathcal{S}}}$
starting with $N=6$ (by contrast, 
$4(N\log_2{(4N)})^{1/2}/d^{1/2+\alpha} < 1$ iff $N \ge 8$).  
Likewise, tighter bounds can be obtained if one is only interested in large 
$N$; for example, one may have
$c=c_N \raw e^{3/4}/\sqrt{2 \pi} \approx 0.84456$ and
$C=C_N \raw e^{1/4}\sqrt{2/\log{2}} \approx 2.1811$, see Appendix E. 
Finally, our methods allow  analyzing separable states on tensor products 
involving spaces  ${\mathbb {C}}^k$ with $k > 2$, leading to non-trivial
but not definitive results; some remarks to that effect are presented in
Appendix I.

\medskip 
Instead of working directly with $\mathcal{D}$ and $\mathcal{S}$,  
we shall consider their  respective symmetrizations
\begin{equation} \label{sym}
\Delta := \conv (-\mathcal{D} \cup \mathcal{D}),  
\ \  \Sigma :=  \conv (-\mathcal{S} \cup \mathcal{S}),
\end{equation}
where all sets are thought of as being contained in the real 
$d^2$-dimensional vector space of self-adjoint elements of
$\calb(\calh)$ (further identifiable with $\mathcal{M}_d^{sa}$,
the space of $d \times d$ complex Hermitian matrices). 
We do that because, firstly, the geometry of {\em symmetric}
convex sets is much better understood than that of the general ones
and, secondly, the specific symmetric sets $\Delta$ and $\Sigma$ 
are familiar objects in geometry of Banach spaces, which allows us 
to refer to known concepts and results. In Appendix D we indicate
how one can treat directly $\mathcal{D}$ and $\mathcal{S}$ without passing to
symmetrizations; however, this yields only a very small improvement in 
the constants $c, C$ in (\ref{vr})
at the price of obscuring somewhat the argument.

It is readily verified that $\Delta$ consists exactly of those (Hermitian) 
elements of $\calb(\calh)$ whose {\em trace class norm} is $\le 1$.
Equivalently, $\Delta$ is the unit ball of the space 
$\mathcal{C}_1^d :=(\mathcal{M}_d^{sa}, \|\cdot\|_1)$, where, for 
$p \in [1, \infty)$,  $\|A\|_p := (\tr (A^*A)^{p/2})^{1/p}$ 
is the Schatten-von Neumann  $p$-norm of the matrix $A$ and
$\mathcal{M}_d^{sa}$ stands for the space of $d \times d$ Hermitian matrices.  
A similar argument shows that
$\Sigma$ is the unit ball of the $N$th  projective tensor power of
$\mathcal{C}_1^2$ (in the sense of the  Banach space theory, see Appendix B). 
We shall denote the corresponding norm
on   $\mathcal{M}_d^{sa}$ by $\|\cdot\|_\pi$.
For future reference, we note that in the above notation $\|\cdot\|_\infty$
corresponds to
$\|\cdot\|_{op}$, the usual norm of a matrix as an operator on the Euclidean
space. We also point out that while in this note we focus on 
($\mathbb {R}$-linear) spaces 
of Hermitian matrices and operators, the
Schatten-von Neumann classes
$\mathcal{C}_p^d$ are most often defined in the literature to include all
(be it real or complex) scalar matrices and not just the self-adjoint ones. 

\medskip The plan of the rest of the argument is as follows. 
First,  using classical general results from convexity, we relate the
volumes  of $\Delta$ and $\Sigma$ to those of $\mathcal{D}$ and $\mathcal{S}$.  
Next, we obtain two-sided
estimates for $\vol \Delta$ and $\vol \Sigma$, which are most
conveniently described using the following concept:  if $K$ is a subset 
of an $n$-dimensional Euclidean space with the unit ball $B$,  we call 
$({\rm vol} \, K/{{\rm vol} \, B})^{1/n}$ the {\em volume radius} of $K$.  
[As hinted earlier, in the present context 
the Euclidean structure is determined by the $2$-norm defined above, 
also often called the Hilbert-Schmidt norm 
or the Frobenius norm, and the inner product is 
$\langle u, v \rangle = \tr\, uv$.]
Equivalently, the volume radius of $K$ is the radius of a Euclidean ball whose
volume is equal to that of $K$. Our approach will determine the volume radius
of 
$\Sigma$ up to a factor which is a power of $\log d$, in particular precisely
on the scale of powers of $d$; this is the principal result of this note. 
The corresponding problem for $\Delta$, the unit ball in the trace class norm, 
is much better understood.  Indeed, two-sided estimates for the volume radius of
$\Delta$ involving a rather large (but universal, i.e., independent of $N$) 
constant follow from an early paper \cite{ST}. Moreover, explicit formulae 
for the volume of $\mathcal{D}$ involving multiple integrals can be produced; 
see \cite{SR} for an analysis of a closely related problem, which can be
routinely modified to yield similar expressions for $\mathcal{D}$.
After a preliminary version of the present note has been circulated, the
author has learned that this circle of ideas has led to a closed formula for the
volume of $\mathcal{D}$ in a very recent work \cite{ZS}; see Appendix E for
more details and \cite{SZ} for related results concerning the Bures
volume. [Undoubtedly, formulae for the volume of $\Delta$ may be similarly
obtained.] The unified argument for estimating the volume radii that is
presented in this paper allows to deduce (from known facts) the value of the
volume radius of $\mathcal{D}$ up to a factor of 2.

\medskip
For the first point, i.e., comparing the volumes of convex sets and 
their symmetrizations, we use a  1958 result of Rogers and Shephard 
(see Appendix C for more details and background) to deduce that 
\begin{equation} \label{SDelta}
\frac{2}{\sqrt{d}} \ \vol \mathcal{D}  \le 
\vol\Delta  \le \frac{2}{\sqrt{d}}  \ 
 \frac{2^{n}}{n+1}\vol \mathcal{D} ,
\end{equation}
where $n = \dim \mathcal{D} = d^2-1$.  [The factor $2/\sqrt{d}$ appears because
it is the distance between  the hyperplanes containing $\mathcal{D}$ and
$-\mathcal{D}$; note that strictly speaking we should be writing
$\vol_n\mathcal{D}$  and $\vol_{n+1}\Delta$ to refer to $n$ and
$n+1$-dimensional volume respectively.]  Similarly 
\begin{equation} \label{SSigma}
\frac{2}{\sqrt{d}}  \ \vol \, \mathcal{S}  \le 
\vol \, \Sigma  \le  \frac{2}{\sqrt{d}}  \ 
 \frac{2^{n}}{n+1} \vol \, \mathcal{S} .
\end{equation}
Combining (\ref{SDelta}) and (\ref{SSigma}) we obtain 
\begin{equation} \label{symmetrization}
 \left(\frac{2^{n}}{n+1}\right)^{-1} \frac{{\vol} \, \Sigma }{{\vol}\,
\Delta}  \ \le \ 
\frac{{\vol} \, \mathcal{S}}{{\vol}\, \mathcal{D}} 
\ \le \ \frac{2^{n}}{n+1} \frac{{\vol} \, \Sigma }{{\vol}\, \Delta} .
\end{equation}
Given that the proper homogeneity is achieved by raising the 
volume ratios to the power $1/n$ (or $1/(n+1)$), we see that one may replace 
$\mathcal{D}$ and $\mathcal{S}$ in (\ref{vr}) by 
$\Delta$ and $\Sigma$ with the accuracy 
of the estimates affected at most by a factor of $2$. 

It remains to estimate $\vol\, \Delta$ and $\vol \, \Sigma $; 
this will be accomplished by comparing each of these bodies 
with the $d^2$-dimensional Euclidean ball $B_{HS}$ (the unit ball 
\wrt the Hilbert-Schmidt norm; we shall also denote by $S_{HS}$ 
the corresponding $d^2-1$-dimensional sphere).  

Concerning $\Delta$, we claim that its volume radius satisfies
\begin{equation} \label{vrDelta}
1/\sqrt{d} \le (\vol \Delta/ \vol B_{HS})^{1/d^2} \le 2/\sqrt{d}
\end{equation}
To show this, we note first the ``trivial" inclusions 
$B_{HS}/\sqrt{d} \subset \Delta \subset B_{HS}$, which just reflect the 
inequalities $\|\cdot\|_2 \le \|\cdot\|_1 \le \sqrt{d} \, \|\cdot\|_2 $ 
between the trace class and the Hilbert-Schmidt norms. The first inclusion
implies  the lower estimate on the volume radius in (\ref{vrDelta}). 
The upper bound is less obvious, but it may be shown by the 
following rather general argument. 
The first step is the classical Urysohn inequality, which in our 
context asserts that
\begin{equation} \label{urysohn}
\left(\frac{\vol \, \Delta}{\vol B_{HS}}\right)^{1/d^2} \le \int_{S_{HS}}
\|A\|_{op} dA \ =: \mu_d,
\end{equation}
where the integration is performed with respect to the normalized 
Lebesgue measure on the Hilbert-Schmidt sphere.  
(For clarity and to indicate flexibility of the approach we shall present  
a general statement and a short proof in Appendix A.) The quantity $\mu_d$ is
most easily  handled by passing to an integral with respect to the standard
Gaussian measure, which reduces the problem to finding expected value of the 
norm of the random Gaussian matrix $G = G(\omega) \in
\mathcal{M}_d^{sa}$, usually  called the Gaussian Unitary Ensemble or GUE.  It
is well known that 
$\mathbb{E} \|G\|_{op} = \gamma_{d^2} \mu_d$, where $\gamma_k
:=\sqrt{2}\Gamma(\frac{k+1}2)/\Gamma(\frac k2)$ for $k \in \mathbb{N}$
(this uses just $1$-homogeneity of the norm),
and it is easy to check  that  
$\sqrt{k-1} < \gamma_k< \sqrt{k}$ for all $k$. 
In other words, 
$\mu_d \sim \mathbb{E} \|G\|_{op}/d$ for large $d$.  On the other hand, 
it is a well-known strengthening of Wigner's semicircle law that 
$\mathbb{E} \|G\|_{op}/\sqrt{d} \raw 2$ as $d \raw \infty$. 
This shows the second inequality in (\ref{vrDelta}) with $2$ replaced 
by  $2+o(1)$. 
We sketch the argument that gives the exact number $2$ in Appendix F
(it follows from known facts, but appears to have been overlooked 
in the random matrix literature), 
yet we will not dwell on it as it intervenes only in the lower estimate 
in (\ref{vr}) and, in any case, the constants in our final 
results are not meant to be optimal.  Indeed, (\ref{vrDelta}) 
combined with (\ref{SDelta}) implies that the volume radius of $\mathcal{D}$ 
is between $\frac12 d^{-1/2}$  and $2d^{-1/2}$, while the formulae
from 
\cite{ZS} allow to deduce that it is equivalent to ${\frac{1}{{e}^{1/4}}}
d^{-1/2}$  as $d \rightarrow \infty$.

\bigskip  We now pass to the analysis of the the volume radius 
of $\Sigma$.  We shall show that
\begin{equation} \label{vrSigma}
1/d^{1+\alpha} \le (\vol \, \Sigma/ \vol B_{HS})^{1/d^2} 
\le C\sqrt{\log{d} \log \log {d}}/d^{1+\alpha} ,
\end{equation}
where $\alpha$ is the same as in (\ref{vr}).  Our main result (\ref{vr})
follows then by combining (\ref{vrDelta}), (\ref{vrSigma}) and
(\ref{symmetrization}). [To be precise, one obtains  {\em a priori }
$1/d^2$ in the exponent, but a more careful analysis of lower order factors
such as $2/\sqrt{d}$ and $1/(n+1)$ appearing in
(\ref{SDelta})-(\ref{symmetrization}) allows to replace
$d^2$ by $\dim \, \mathcal{S}  = d^2-1$ without any loss in the constants.]  

Before proceeding, let us compare  (\ref{vrSigma})  with the results of 
\cite{GB}, which estimate from {\em below} the in-radius of 
$\mathcal{S}$ in the Hilbert-Schmidt metric by a quantity that is of
order of $d^{-\eta}$, where $\eta = 3/2$. The easy {\em upper } bound on that
radius is the in-radius of $\mathcal{D}$, which equals $1/\sqrt{d(d-1)} =
O(d^{-1})$. The second inequality in (\ref{vrSigma}) yields (for large $N$) a
better upper estimate that roughly corresponds to $\eta = 1+ \alpha \approx
1.094361$; we elaborate on these and related issues in Appendix~H.   

\medskip
The first step towards showing (\ref{vrSigma}) will be to replace the sets
$\Sigma_N$ by  their affine images which are more ``balanced;" 
this will also explain the appearance of the mysterious number $\alpha$ 
in the exponents. 

Consider first the sets in question when $N=1$. 
As is well known, $\mathcal{S}_1$ and $\mathcal{D}_1$ 
both coincide with the Bloch ``ball," which geometrically is a 
(solid) Euclidean ball of radius $1/\sqrt{2}$, the boundary of which is 
the Bloch sphere $\mathcal{T}_1$ consisting of pure states on 
$\calb(\mathbb{C}^2)$  
(further identifiable with rank $1$ projections on $\mathbb{C}^2$). 
Accordingly,  $\Sigma_1 = \Delta_1$
is  a $4$-dimensional cylinder whose base is the Bloch ball and whose 
axis is the segment $[-I_2/2, I_2/2]$ of Euclidean length $\sqrt{2}$.
For definiteness, let us identify $\mathcal{M}_2^{sa}$ with $\mathbb{R}^4$
via the usual basis 
$\{I_2/\sqrt{2}, \sigma_x/\sqrt{2}, \sigma_y/\sqrt{2}, \sigma_z/\sqrt{2}\}$,
where $\sigma_x, \sigma_y$ and $\sigma_z$ are the Pauli matrices  
(the factors $1/\sqrt{2}$ make this basis orthonormal in the 
Hilbert-Schmidt sense). Let now $A$ be a linear map on $\mathcal{M}_2^{sa}$
which is diagonal in that basis and whose action is defined by 
$AI_2 = I_2/\sqrt{2}$, $A\sigma_i = \sqrt{3/2}\, \sigma_i$ for $i=x,y,z$.
Set  $\tilde{\Sigma}_1 := A \Sigma_1$;  
the important properties of $A$ and $\tilde{\Sigma}_1$ are
 
\medskip \noindent (i) \ the image of the Bloch sphere 
$A \mathcal{T}_1 =: \tilde{\mathcal{T}}_1$ is geometrically a 
$2$-dimensional sphere of radius $\sqrt{3}/2$ centered at 
$I_2/\sqrt{8}$
and, as the Bloch sphere itself, it is contained in the 
{\em unit } Euclidean sphere $S_{HS}$; this implies that 
$\tilde{\Sigma}_1 \subset B_{HS}$

\smallskip \noindent (ii) \, $\det A =\sqrt{27/16}$ and so 
$\vol \, \tilde{\Sigma}_1 = \sqrt{27/16} \; \vol \, \Sigma_1$

\smallskip \noindent (iii) vertices of any regular tetrahedron 
inscribed in  $\tilde{\mathcal{T}}_1$ form an orthonormal basis
in $\mathcal{M}_2^{sa}$.

\medskip \noindent 
The geometric property of the set $\tilde{\Sigma}_1$, 
which arguably is the reason for its relevance, 
is that the ellipsoid of smallest volume containing it (the so-called
L\"owner ellipsoid of $\tilde{\Sigma}_1$) is the Euclidean ball. 
An equivalent and perhaps more natural
point of view would be to compare 
$\Sigma_1$ with its own L\"owner ellipsoid. This is in turn equivalent 
to replacing the Hilbert-Schmidt inner product 
$\langle u, v\rangle = \tr uv$ with 
\begin{equation} \label{killing}
\tr ((Au)(Av)) = (3\, \tr uv - \tr u \,\tr v)/ 2 .
\end{equation}  
It is likely that  this non-isotropic inner product and
objects associated with it play an important role in the theory.
In particular, we obtain this way ellipsoids which -- from the volumetric point
of view -- are nearly non-distinguishable from  $\mathcal{S}$ or $\Sigma$,
and which still enjoy  certain permanence relations with respect to the action
of the unitary group.

If $N > 1$, we set $\tilde{\Sigma}=\tilde{\Sigma}_N := A^{\otimes N}
\Sigma_N$.  Since 
$\det A^{\otimes N} = (\det A)^{N \cdot 4^{N-1}} = ((27/16)^{N/8})^{d^2}
=(2^{\alpha N})^{d^2}= (d^\alpha)^{d^2}$, 
we deduce that (\ref{vrSigma}) is equivalent to
\begin{equation} \label{vrSigmamod}
1/d \le (\vol \, \tilde{\Sigma}/ \vol B_{HS})^{1/d^2} 
\le C\sqrt{\log{d} \log \log {d}}/d ,
\end{equation}
For the lower estimate in (\ref{vrSigmamod}) we shall produce a simple  
(and seemingly not very optimal) geometric argument. 
Let $u_1, u_2, u_3, u_4$ be vertices of  any regular
tetrahedron  inscribed in  $\tilde{\mathcal{T}}_1$. By the property 
(iii) above, $(u_j)_{j=1}^4$ is an orthonormal basis of $\mathcal{M}_2^{sa}$.
Accordingly, the set 
$\tilde{U} := \{ u_{j_1} \otimes u_{j_2} \otimes \ldots \otimes u_{j_N} \}$, 
where each $j_i$ ranges over $\{1,2,3,4\}$ is an orthonormal basis of 
$\mathcal{M}_d^{sa}$.  The first inequality in (\ref{vrSigmamod})
follows now from the inclusions $\tilde{U}, -\tilde{U} \subset
\tilde{\Sigma}_N$ and 
$B_{HS}/d  \, \subset  \conv (-\tilde{U} \cup \tilde{U})$ 
(the latter is a consequence of the orthogonality of elements of $\tilde{U}$).

The above argument may appear rather ad hoc, and so it may be instructive 
to rephrase it in the language of geometry of Banach spaces. 
Let $A_1$ be a linear map from $\mathbb{R}^4$ to $\mathcal{M}_2^{sa}$ 
which sends the standard unit vector basis onto vertices of any regular
tetrahedron  inscribed in  $\mathcal{T}_1$. By construction,
$A_1$ is a contraction from $\ell_1^4$ to $\mathcal{C}_1^2$  
and so its $N$-th tensor power $A_1^{\otimes N}$  induces a
contraction  between the respective projective tensor powers of $\ell_1^4$ 
and $\mathcal{C}_1^2$ (where $\ell_1^k$ denotes $\mathbb{R}^k$ 
endowed with the norm $\|(x_j)\| = \sum |x_j|$). As the projective tensor 
product of $\ell_1$-spaces is again an $\ell_1$-space, it follows that 
$\Sigma_N$ contains the image under $A_1^{\otimes N}$ 
of the unit ball of $\ell_1^{d^2}$, and hence the image of the 
Euclidean ball of radius $1/d$. In particular,
$\vol \, \Sigma/\vol (A_1^{\otimes N} B_{HS}) \ge (1 /d)^{d^2}$. 
On the other hand, one verifies (directly, or by noticing that 
$A = |A_1^{-1}| = (A_1^{-1*}A_1^{-1})^{1/2}$) that
$\vol (A_1^{\otimes N} B_{HS}) = ((16/27)^{N/8})^{d^2}\vol B_{HS}$, 
which substituted into the preceding estimate gives exactly the first inequality
in (\ref{vrSigmamod}). 

We note that using for the above calculations the larger volume of 
the image of the $\ell_1^{d^2}$ ball (resp., $\conv (-\tilde{U} \cup
\tilde{U})$) would only result in a slightly better constant
$c$ in (\ref{vr}) (the value $c=\sqrt{e/8\pi}$
that was mentioned earlier). This is because the volume radius of the unit ball
in
$\ell_1^{m}$ is roughly the same as that  of the inscribed Euclidean ball, the
ratio between the two  is $\sqrt{2e/\pi} \; (1-O(1/m))$. This property is behind
many striking phenomena discovered in the asymptotic theory of finite 
dimensional normed spaces, and is closely related to 
our upper estimates for $\vol \, \Sigma$ and $\vol \, \mathcal{S}$, 
to which we pass now.  

To prove the upper estimate in (\ref{vrSigmamod}), we shall again use the
Urysohn inequality. 
Analogously to (\ref{urysohn}) and to the reasoning that followed it, we get
\begin{equation} \label{mstarpi}
\left(\frac{\vol \tilde{\Sigma}}{ \vol B_{HS}} \right)^{1/d^2} \le
\int_{S_{HS}} \max_{X \in \tilde{\Sigma}} \, \tr(XA)  \, dA 
= \gamma_{d^2}^{-1} \mathbb{E} \max_{X \in \tilde{\Sigma}} \, \tr(XG) 
\end{equation}
and so it remains to show that
the above expectation is $O(\sqrt{\log{d} \log \log {d}})$.   The expression
under the expectation can be thought of as a maximum of a Gaussian process
indexed by $\tilde{\Sigma}$  (this just means the family 
$\tr(XG(\omega))$,  $X \in \tilde{\Sigma}$, of jointly
Gaussian random  variables). 
There are several methods of differing sophistication which
can be used to estimate  the expectation of such a maximum. The two leading
ones are the Fernique-Talagrand majorizing measure theorem,
which gives the correct asymptotic order, but is usually difficult to apply, and
the Dudley majoration (by the metric entropy integral), which is almost 
as precise and usually easier to handle;  see \cite{L} for a comprehensive
exposition. We shall employ here an even simpler ``one-level-discretization"
method   which, in our context, yields approximately the same result as the
Dudley  majoration, and which we now describe in elementary language.  

Let $\mu$ be the standard
Gaussian measure on $\mathbb{R}^m$ (i.e., the one given by the density 
$m(x)=(2\pi)^{-m/2} \exp(-|x|^2/2)$, where $|\cdot |$ is the corresponding
Euclidean norm) and let 
$F \subset \mathbb{R}^m$ be a finite set contained in a ball of radius $R$. 
Then 
\begin{equation} \label{mstardiscrete}
\int_{\mathbb{R}^m} \max_{y \in F} \, \langle y,x \rangle \, d \mu (x)
\le  R \sqrt{2 \log (\# F)} ,
\end{equation}
where $\#$ stands for the cardinality of a set. The estimate above is usually
quoted with a different numerical constant appearing in place of $2$, 
but it is not difficult -- even if somewhat tedious -- to verify that it holds 
in the form stated above.  The idea now is to construct a finite set 
$F \subset \tilde{\Sigma}$ such that $\conv F \supset r \tilde{\Sigma}$ for an
appropriate 
$r \in (0,1)$; it will then follow that 
\begin{equation} \label{mstarr}
\mathbb{E} \max_{X \in \tilde{\Sigma}} \, \tr(XG) 
\le r^{-1} \sqrt{2 \log (\# F)} .
\end{equation}
[We note that the maxima of the type appearing in (\ref{mstarpi}), 
(\ref{mstardiscrete}) or (\ref{mstarr}) do not change if we  replace 
the underlying (closed) set $F$ 
by its convex hull or, conversely, by its extreme points.] 
Specifically, $F$ will be  a ``sufficiently dense" subset of the set
of extreme points of $\tilde{\Sigma}$, i.e., of 
$-\tilde{\mathcal{T}} \cup \tilde{\mathcal{T}}$,  where
$$
\tilde{\mathcal{T}} = \tilde{\mathcal{T}}_N = 
\{A\rho_1 \otimes A\rho_2 \otimes \ldots \otimes A\rho_N  \}  
$$ 
and where each $\rho_j$ is a pure state on $\calb(\mathbb{C}^2)$  
(i.e., an element of the Bloch sphere). In other words, 
$\tilde{\mathcal{T}}$ is a tensor product of $N$ copies of
$\tilde{\mathcal{T}}_1=B \mathcal{T}_1$ 
which, as we noted earlier, is geometrically a $2$-dimensional sphere 
of radius 
$\sqrt{3}/2$ contained in the unit sphere of the $4$-dimensional 
Euclidean space. 

We start by constructing an appropriate dense subset 
(usually called {\em net}) in each copy of 
$\tilde{\mathcal{T}}_1$ and then consider tensor products of those nets.
To facilitate references to existing literature we first look at the 
{\em unit} Euclidean ball $S^2$ rather than $\tilde{\mathcal{T}}_1$.
Let $\delta \in (0,\sqrt{2})$ and let $\mathcal{N}$ be a $\delta$-net 
of $S^2$, i.e., a subset such that the union of balls of radius $\delta$
centered  at points of $\mathcal{N}$ covers $S^2$. An elementary
argument shows that $\conv \mathcal{N}$ contains then a  ball of radius
$(1-\delta^2/2)$ centered at the origin.  If now 
$F_1 \subset \tilde{\mathcal{T}}_1$ is an appropriate dilation of 
$\mathcal{N}$ (i.e., with the ratio $\sqrt{3}/2$), then  
$\conv F_1$ contains a  ball  
of radius $(1-\delta^2/2)\cdot \sqrt{3}/2$  
(in the $3$-dimensional affine space containing $\tilde{\mathcal{T}}_1$)
with the same center as that of of $\tilde{\mathcal{T}}_1$.
It follows that 
$\conv (-F_1 \cup F_1) \supset (1-\delta^2/2) \, \tilde{\Sigma}_1$
and consequently if we set $F: = (-F_1 \cup F_1)^{\otimes N} 
= F_1^{\otimes N} \cup (- F_1^{\otimes N})$,  
then $\conv F \supset (1-\delta^2/2)^N \tilde{\Sigma}_N$. 

In remains to find a reasonable bound on  $\# F = 2(\# F_1)^N = 2(\#
\mathcal{N})^N$. A standard argument comparing 
areas of caps and that of the entire sphere shows that one
may have a $\delta$-net of $S^2$ of cardinality 
$< 16/\delta^2$. 
[This bound is not optimal; the asymptotically -- as $\delta \raw 0$ --
 correct order for cardinalities of efficient $\delta$-nets of $S^2$ is  
$(2/\sqrt{3})^3\pi/\delta^2$, but coverings of
Euclidean spheres, even in  dimension $2$, do not appear to be 
completely understood.]
This leads to an estimate 
$\#F < 2(16/\delta^2)^{N}$, 
which in combination with (\ref{mstarr}) gives
\begin{equation} \label{2NlogN}
\mathbb{E} \max_{X \in \tilde{\Sigma}} \, \tr(XG) \le
(1-\delta^2/2)^{-N} \sqrt{2 \log (2(16/\delta^2)^{N})} .
\end{equation}
Optimizing the expression on the right hand side over $\delta \in (0,
\sqrt{2})$  yields a quantity that is of order $\sqrt{2 N \log{N}}$ 
for large $N$ (choose, for example, $\delta = (N \log{N})^{-1/2}$),
as required to complete the proof of (\ref{vrSigma}) (and hence of (\ref{vr})). 
Moreover, substituting the obtained 
bound into (\ref{mstarpi}) and verifying numerically small values of $N$  yields
\begin{equation} \label{mstarfinal}
\left(\frac{\vol \tilde{\Sigma}}{ \vol B_{HS}}\right)^{1/d^2}\le \frac{\sqrt{4N
\; \log_2{(4N)}}}{d} = \frac{\sqrt{4\log_2{d} \; \log_2{(4\log_2{d})}}}{d}
\end{equation}
(note that the inequality is trivial for $N=2$), which implies that (\ref{vr})
holds with the third member  of the form $\frac{4 \sqrt{N \log_2{(4N)}}}{d^{1/2
+
\alpha}}$. This may be somewhat improved for small to moderate values of $N$ by
using estimates on cardinalities of nets of $S^2$ listed in \cite{HSS}, see
Appendix G. 

\vskip.25cm
{\small {
\noindent {\bf Acknowledgments.}   This research has been partially supported
by  a grant from the National Science Foundation (U.S.A.). 
The final part of the research has been
performed and the article written up while the author visited Stefan Banach
Centre of Excellence at the Institute of Mathematics of the Polish Academy of
Sciences in the summer of 2003.
 The author thanks Dorit
Aharonov whose inquiry initiated this project, and who subsequently offered
various helpful remarks, and Vitali Milman who publicized the inquiry in the
``asymptotic geometric analysis" community.
He further thanks G. Aubrun, L. Gurvits, M. Lewenstein, P. Slater and K.
\.Zyczkowski who commented on the preliminary version of the paper.
}}

\vskip.25cm
\medskip \noindent  {\bf Appendix } {\bf A }{\em The
Urysohn inequality. } If $K$ is a convex body in the
$m$-dimensional  Euclidean space which contains $0$ in its interior, then 
$$
\left( \frac{\vol K}{\vol B} \right)^{1/m}= 
\left( \int_{S^{m-1}} \|x\|_K^{-m} dx \right)^{1/m} 
\ge \int_{S^{m-1}} \|x\|_K^{-1} dx 
\ge \left( \int_{S^{m-1}} \|x\|_K dx \right)^{-1},
$$
where $\|x\|_K$ is the gauge of $K$ (the norm for which $K$ is the 
unit ball if $K$ is $0$-symmetric -- which is the case in the main text) 
and  the integration is performed with respect to the normalized Lebesgue 
measure on the sphere $S^{m-1}$. If $K$ is $0$-symmetric, this may be combined 
with the Santal\'o inequality \cite{santalo} which asserts that
$$
\frac{\vol K}{\vol B} \cdot \frac{\vol K^\circ}{\vol B} \ \le \ 1,
$$ 
where $K^\circ := \{x : \langle x, y \rangle \le 1 \ 
{\rm for \; all } \ y \in K \}$ is the polar of $K$, 
to obtain $(\vol K^\circ/\vol B)^{1/m} \le \int_{S^{m-1}} \|x\|_K dx$, and the 
Urysohn's inequality 
$$
(\vol K/\vol B)^{1/m} \le \int_{S^{m-1}} \|x\|_{K^\circ} dx =
\int_{S^{m-1}} \max _{y \in K}{\langle x, y \rangle} dx 
$$
follows by exchanging the roles of $K$ and $K^\circ$.
Moreover, since the Santal\'o inequality holds also for 
not-necessarily-symmetric sets after an appropriate translation, 
the above inequality holds for any (measurable) bounded set $K$.
Indeed, the integral on the right equals $1/2$ of the {\em mean
width } of $K$, a well-known classical geometric parameter of a set in the
Euclidean space, which does not change if
$K$ is replaced  by its translation. It is primarily the mean width of various
sets -- and not directly volume -- that is being majorized throughout this
paper.

This is not the most elementary proof of the Urysohn inequality, 
but one that offers a lot of flexibility. 
For example, the repeated applications of the H\"older inequality
in the first chain of inequalities above 
can be modified to yield as the last expression $\left( \int_{S^{m-1}}
\|x\|_K^p dx
\right)^{-1/p}$ for an arbitrary $p >0$ and, letting $p \raw 0$,  
the geometric mean  
$\exp{\left( - \int_{S^{m-1}} \log {\|x\|_K }dx \right)}$.  
Similar inequalities also hold if $p \in [-n,0)$, and the case $p = -n$ 
is of course the strongest statement of such nature, 
the Santal\'o inequality itself. We also take this opportunity to point 
out that the Santal\'o inequality and the so-called reverse Santal\'o
inequality \cite{BM} together imply that the volume radius of a convex set and
its polar are roughly (i.e., up to universal multiplicative constants)
reciprocal.

The application of the Urysohn inequality in (\ref{urysohn}) uses
implicitly the elementary fact that  $\|\cdot \|_{\Delta^\circ} = \| \cdot
\|_{op}$. In other words, the trace class norm $\| \cdot \|_1$ and the operator
norm $\| \cdot \|_{op}$ are dual with respect to the trace duality. For the
sets $\Delta_N$, the Urysohn inequality gives the correct order of the
volume radius, cf. (\ref{vrDelta}). However, this is not always the case, even
for rather regular convex bodies.  For example, if
$K$ is the unit ball of 
$\ell_1^m$, then its volume radius  -- as we have already mentioned --  
exceeds the radius of the inscribed Euclidean ball by less than
$\sqrt{2e/\pi}$, while the upper bound obtained from the Urysohn inequality 
contains a parasitic factor which is of order $\sqrt{\log {m}}$. It is thus
conceivable that the logarithmic factors in (\ref{vr}) can be replaced by
universal numerical constants. On the other hand, 
if we {\em do} use the Urysohn inequality to establish an upper bound for the
volume radius of $\tilde{\Sigma}$ (cf. (\ref{vrSigmamod}),
(\ref{mstarpi}), (\ref{mstarfinal})), then our estimates can not be
substantially improved.  Indeed, since our argument showed that $\tilde{\Sigma}$
contained a rotation of the unit ball of $\ell_1^{d^2}$, the previous remark 
implies that the $\sqrt{N}=\sqrt{\log_2 d}$ factor can not be then avoided,
and so it is only the $\sqrt{\log \log d}$ factor that can possibly be
eliminated by more careful majorizing of $\mathbb{E}
\|G\|_{\tilde{\Sigma}^\circ}$. 

Finally, we mention that there exist
general volume estimates for convex hulls of finite sets which are
asymptotically more precise than the one we derive from the Urysohn inequality
(see \cite{ball} and its references). However, these estimates are equivalent
to the ones presented here in the relevant range of parameters and, moreover,
their formulations available in the literature contain unspecified numerical
constants, which would make the corresponding bounds difficult to apply for
specific values of $N$.

\medskip \noindent  {\bf Appendix} {\bf B }{\em 
Projective tensor products of normed spaces.} If $X$ and $Y$ are 
(say, finite dimensional) normed spaces, their tensor product $X \otimes Y$ may
be endowed with the  projective tensor product norm $\| \cdot \|_\pi$ defined by

$ \qquad \quad
\| \tau \|_\pi := \inf \left\{ \sum_{j=1}^m \|x_j\| \cdot \|y_j\| \ : \ 
\sum_{j=1}^m x_j \otimes y_j = \tau \right\} .
$

\noindent The resulting normed space is usually denoted $X \hat{\otimes} Y$ or  
$X \otimes_\pi Y$.
If $B_X$ and $B_Y$ are unit balls of 
$X$ and $Y$ respectively, it follows that the unit ball of
$X \hat{\otimes} Y$ coincides with 

$\conv \left\{ x \otimes y  : x \in B_X, y \in B_Y \right\} = \conv \left\{ x
\otimes y  : x
\in {\rm ext}B_X, y \in {\rm ext}B_Y \right\}$,

\noindent where ${\rm ext}(K)$ denotes the set of extreme points of $K$.
If $X=Y=\mathcal{C}_1^2$, the analysis is further simplified by the fact 
that the set of extreme points of $\Delta_1$, the unit ball of
$\mathcal{C}_1^2$, is of the form   $-\mathcal{T}_1 \cup \mathcal{T}_1$,
where $\mathcal{T}_1$ is the set of pure states. 
The fact that the set $\Sigma_2$ is the unit ball of  
$\mathcal{C}_1^2 \hat{\otimes} \mathcal{C}_1^2$ follows directly from these
identifications.  
Projective tensor products of more than two spaces are
defined analogously (or by induction), and one similarly checks that 
the unit ball of the $N$th projective tensor power of
$\mathcal{C}_1^2$ is $\Sigma_N$.  Tensor products involving spaces 
${\mathbb {C}}^k$ with $k > 2$ may be treated in the same way. For example,
the symmetrization of the set of separable states on 
$\calb({\mathbb {C}}^{k_1}\otimes {\mathbb {C}}^{k_2}\otimes 
\ldots {\mathbb {C}}^{k_m})$ can be identified with the unit ball in 
$\mathcal{C}_1^{k_1}\hat{\otimes}\mathcal{C}_1^{k_2} \hat{\otimes} \ldots
\hat{\otimes}\mathcal{C}_1^{k_m}$. The problem of the relative size of the 
set of separable states on $\calb(({\mathbb {C}}^{D})^{\otimes N}$, or $N$
qudits, was investigated in \cite{qudits}.  While a more definitive treatment
of the higher-dimensional case will be presented elsewhere, we offer some
comments on the topic in Appendix I.

On the more elementary level, the projective tensor square of a Euclidean space
$\mathbb{C}^{k}\hat{\otimes}\mathbb{C}^k$ can be identified with 
$(\calm_k, \| \cdot \|_1)$ and the contractively complemented subspace of its
self-adjoint elements is, in our notation, $\mathcal{C}_1^k$.  We also mention
in passing that  the dual space to $X \hat{\otimes} Y$ can be identified with
the so-called {\em injective} tensor product of the duals $X^*$ and $Y^*$, and
so
$\Sigma^\circ$ can be thought of as a unit ball in the  injective tensor power
of the Hermitian part of
$\calb(\mathbb{C}^2)$. While  making this identification explicit doesn't seem
to help our analysis at the present level of depth, we mention in passing that
various existing criteria for detecting entanglement use separation theorems
for convex sets which are based on a form of this duality.

\medskip \noindent {\bf Appendix} {\bf C} {\em The Rogers-Shephard results
on symmetrizations of convex sets.}  Let $W \subset \mathbb{R}^{n+1}$ be an
$n$-dimensional convex set and denote by $h$ the distance from the affine
hyperplane $H$ spanned by $W$ to the origin. Let $\Omega$ be the
symmetrization of $W$, i.e., $\Omega := \conv (-W \cup W)$. It was shown in 
\cite{RS2} that then 
\begin{equation} \label{rs}
2h \ \vol W  \le 
\vol\Omega  \le 2h  \ 
 \frac{2^{n}}{n+1}\vol W ,
\end{equation}
where by $\vol W$ and $\vol\Omega$ we mean the $n$- and the $n+1$-dimensional
volume respectively. To explain the factors appearing in (\ref{rs}) we note
that the inequalities becomes equalities  if $W$ is centrally symmetric for 
the first one (this is simple) and if $W$ is a simplex for the second (this is
the heart of the Rogers-Shephard result). 

To further clarify 
the first inequality in (\ref{rs}) (used in the upper estimates on the volume of
separable states, which is the main point of this note) we point out that it is
actually a simple consequence of a much older theorem  of Brunn-Minkowski, and
more specifically of the following  corollary of that theorem.  

\smallskip {\em Let } $K$ {\em be an }
$n+1$-{\em dimensional convex body, }
$u$ {\em -- a vector  in the ambient space containing } $K$ {\em and } $H$
{\em  -- a
hyperplane in  that space. Then the function } $t \raw \vol(K \cap
(tu+H))^{1/n}$  {\em (the }$n${\em -dimensional volume) is concave on its
support.}

\smallskip
If we apply the above fact with  $K = \Omega$ and $u$ - a unit vector
perpendicular to $H$, then the function $\phi(t):= \vol(\Omega \cap
(tu+H))$,  being even on $[-h,h]$, must attain its maximum at 0 and minimum at
$h$ and $-h$. The first inequality in (\ref{rs}) follows then from the Cavalieri
principle. A version of the second inequality, which would be sufficient for
our purposes,  follows similarly from  the estimate $\phi(0) \le  2^{-n} {2n
\choose n}\vol W$,  which is the main result of \cite{RS}.

\medskip \noindent {\bf Appendix } {\bf  D }{\em Working directly with 
non-symmetric sets.}  Similar but slightly more complicated arguments may be
used to obtain {\em upper} estimates for the volumes of the non-symmetric sets
$\mathcal{D}$ and $\mathcal{S}$ by studying directly these sets and not their
symmetrizations  $\Sigma$ and $\Delta$. 
[In principle, this could help to avoid the parasitic factors $2^{n}/(n+1)$ -- 
where $n=d^2-1$ -- when passing from $\Sigma, \Delta$  to $\mathcal{S},
\mathcal{D}$.] In both cases it is convenient to pass to a translate of the set
in question obtained by subtracting the appropriate multiple of the maximally
mixed state
$I_d/d$,  and to consider the translates as subsets of $H_0$, the
$d^2-1$-dimensional space of matrices with vanishing trace. 

For the set $\mathcal{D}$ (translated by $I_d/d$), the quantity which replaces
$\| \cdot \|_{op}$ in the analogue of (\ref{urysohn}) is $\lambda_1(\cdot)$, the
largest eigenvalue of a matrix.  This is of course dominated by the norm, and
since the (random Gaussian) trace $0$ matrix $G_0$  can be represented as a
conditional expectation of the general  Gaussian matrix $G$, it follows -- by
the convexity of the norm or of the largest  eigenvalue -- that  
 $\mathbb{E} \,\lambda_1(G_0) \le \mathbb{E} \|G\|_{op} \le 2 \sqrt{d}$ which,
after some work, leads to an upper estimate for the volume radius of
$\mathcal{D}$ identical to that of $\Delta$ obtained in  (\ref{vrDelta}).
[To fully justify the steps above one needs to appeal to Appendices A and F.]

For the set $\mathcal{S}$, we pass first to the face $\tilde{\mathcal{S}}$ of 
the rescaled set $\tilde{\Sigma}$ that corresponds to $\mathcal{S}$, and then
subtract $I_d/d^{3/2}$ (the difference \wrt the case of $\mathcal{D}$ is due to
the rescaling). Next, we ``approximate" the translate by sets built from 
the points corresponding to elements of $F$.  There are several differences
between this setting and that described in the main text,   but they can be
accounted for fairly easily.  The good news is that the new points are not on
the sphere since the component in the direction of $I_d$ was subtracted, 
but this improves our estimate on $\vol \, S$ only by a factor 
$1- O(d^{-1})$. A loss which is even less significant is due to the fact that 
by reducing the dimension by 1 our formulae will involve the quantity  
$\gamma_{d^2-1}$ rather than $\gamma_{d^2}$.  A somewhat more substantial 
loss comes from the fact that due to the rescaling the width of  
$\tilde{\Sigma}$ in the direction of $I_d$ is different from that of 
$\Sigma$ by a factor of $2^{N/2} = d^{1/2}$; this affects the relationships
between volumes of these bodies and those of $\tilde{\mathcal{S}}$
and $\mathcal{S}$,  and consequently our estimates,  by the same factor. 
However,  since we are in dimension $d^2-1$, the loss in the {\em volume 
radius } is a not-so-significant $1+ O(\log{d}/d)$. The final issue that 
needs to be analyzed is that while we knew that  
$\conv F \supset (1-\delta^2/2)^N \tilde{\Sigma}$,  it is not {\em a priori
} clear that a similar inclusion holds for the face
$\tilde{\mathcal{S}}$ (or, more precisely, for its translate
$\tilde{\mathcal{S}} -I_d/d^{3/2}$). While for a general convex set
$K
\subset H_1$ the relationship between $K$ and its symmetrization ${\rm conv}
(-K \cup K)$  may be more involved, using the fact that $\tilde{\mathcal{S}}$
is a convex hull of points contained in a {\em sphere } centered at
$I_d/d^{3/2}$ we can infer that 
$\conv F \supset (1-\kappa) \tilde{\Sigma}$ implies 
$\conv (F \cap \tilde{\mathcal{S}}) \supset I_d/d^{3/2} + (1-2\kappa) 
(\tilde{\mathcal{S}}-I_d/d^{3/2})$, and the difference between the factors
$1-\kappa$  and $1-2 \kappa$ is asymptotically insignificant.

While the above argument allows to avoid the symmetrizations while estimating
the volume radii of $\mathcal{D}$ and $\mathcal{S}$ from above, we get the
same majorants as in (\ref{vrDelta}), (\ref{vrSigma}). Moreover, we still need
to look at the symmetrizations for the lower bounds which are needed
to derive (\ref{vr}). [It is also possible to obtain a lower
bound on the volume radius of $\mathcal{S}$ directly by noticing that
$\mathcal{S}$ contains a simplex spanned by the elements of $(A^{\otimes
N})^{-1} \tilde{U}$, but since for the simplex the relevant Rogers-Shephard
inequalities become equalities, there is again no significant improvement.]

\medskip \noindent {\bf Appendix } {\bf E } {\em The exact expressions on 
the volume of $\cd$ and the constants in } (\ref{vr}) {\em as $N \raw \infty$.} 
After a preliminary version of this note has been  posted, the author has
learned that a closed formula for the volume of
$\mathcal{D}$ was derived in a very recent work \cite{ZS}. While we were able 
to calculate the volume radius of $\mathcal{D}$ to within a factor of
2 by the same methods that were employed to analyze $\cs$ and with
very little extra work, it is instructive to compare the so obtained estimates 
to those that can be deduced from the exact formula, which in our notation
reads 
\begin{equation} \label{closed}
\vol \mathcal{D}(\mathbb{C}^d) = \sqrt{d} \, (2 \pi)^{d(d-1)/2} \ 
\frac{{\prod_{j=1}^d \Gamma(j)}}{\Gamma(d^2)}
\end{equation}
A tedious but routine calculation based on the Stirling formula shows that 
the volume radius of $\mathcal{D}(\mathbb{C}^d)$ behaves as
${\frac{1}{{e}^{1/4}}} d^{-1/2}(1+ O(d^{-1}))$  as $d \rightarrow \infty$ 

We now recall the refinements related to the 
volume radii of $\cs_N$  and $\Sigma_N$ suggested in the main  text. 
First, we had the argument that gave $c=\sqrt{e/8\pi}$ in (\ref{vr}),
based on using the volume of 
the unit ball in $\ell_1^{d^2}$, i.e., $2^{d^2}/(d^2)!$, 
as a lower bound for $\vol \tilde{\Sigma}$  (see the comments following
(\ref{vrSigmamod})). Next, we noted that, for large $N$, the expressions in 
(\ref{2NlogN}) can be majorized by a quantity that is of order   
$\sqrt{2 N \log{N}}$.    Combining these with the improvement related to
$\mathcal{D}_N$ we are led to an asymptotic version of (\ref{vr}) with 
$c_N \raw e^{3/4}/\sqrt{2 \pi} \approx 0.844561$ and
$C_N \raw e^{1/4}\sqrt{2/\log{2}} \approx 2.1811$.

\medskip \noindent {\bf Appendix }  {\bf F } 
{\em Norms of GUE matrices and the constant $2$ in }
(\ref{vrDelta}).  It has been known for some time (in fact in a much more
general setting) that if $G = G(\omega)$ is the random matrix distributed
according to the standard Gaussian measure on $\mathcal{M}_d^{sa}$
(usually  called the Gaussian Unitary Ensemble or GUE), then, for large $d$, 
$\|G\|_{op}$ is, with high probability, close to $2\sqrt{d}$ . We sketch
here a derivation,  from known facts, of the arguably elegant inequality 
$\mathbb{E} \|G\|_{op} < 2\sqrt{d}$, valid for any $d$,
which appears to have been overlooked in the random matrix theory 
literature. 
Similar inequalities are known for Gaussian matrices all 
whose entries are independent or for real symmetric matrices (also known as 
the GOE  ensemble; however, in the latter case the precise inequality seems to
have been established only for the  largest eigenvalue, and not for the norm),
see
\cite{DS}.  Analogous inequalities with the expected value replaced by the
median can probably be deduced -- at least for large $d$ -- from \cite{TW1,
TW2}.

Our starting point are the recurrence formulae for the (even) moments 
$a_p =a_p(d) := d^{-1} \mathbb{E} \, \tr{\left((G/2)^{2p}\right)}$,
$p \in \mathbb{N}$, derived, e.g., in \cite{HT} (see also \cite{ledoux},
formulae  (6) through (9), for a similar argument and a related estimate)
$$
a_p=\frac{2p-1}{2p+2}\left(a_{p-1} +
\frac{p(p-1)}{4d^2} \frac{2p-3}{2p}a_{p-2}\right),
$$
with $a_0=1$ and $a_1=1/4$. From these one easily derives by induction 
$$
a_p \le \frac{1}{2^{2p}(p+1)} {{2p} \choose p} 
\prod_{j=1}^p\left(1+ \frac{j(j-1)}{4d^{2}}\right).
$$
[This estimate is actually asymptotically precise for $p=o(d)$.] 
Next, using successively the Stirling formula to majorize the binomial
coefficient, the inequalities $1+x\le e^x$ and $\sum_{j=1}^pj(j-1) \le
p^3/3$ to estimate the product, and denoting $t=pd^{-2/3}$, we arrive at
$$
\mathbb{E} \, \tr{\left((G/2)^{2p}\right)} = d a_p \le
d \ \frac{e^{p^3/12d^2}}{\sqrt{\pi} \, p^{3/2}} = \frac{e^{t^3/12}}{
\sqrt{\pi} \, t^{3/2}} .
$$
Hence 
$$
\frac 12 \mathbb{E} \|G\|_{op} < \, \left(\mathbb{E} \,
\tr{\left((G/2)^{2p}\right)}\right)^{1/2p} \le 
\left[ \left( \frac{e^{t^3/6}}{
\pi \, t^3}\right)^{1/4t} \right]^{1/d^{2/3}}.
$$
This is valid for $t>0$, at least if the corresponding value of 
$p=td^{2/3}$ is an integer. 
The minimal value of the expression in brackets over
$t>0$ is attained at $t \approx 1.38319$ and is approximately 
$0.738542 \approx \exp(-0.303077) < e^{-0.3}$. Since for sufficiently large
$d$  the interval corresponding to values which are $< e^{-0.3}$ contains 
an element of $d^{-2/3} \mathbb{N}$, we deduce that for such $d$ we have
$\mathbb{E} \|G\|_{op} < 2 e^{-0.3d^{-2/3}}$.  A more careful checking
shows that in fact the inequality $\mathbb{E} \|G\|_{op} < 2 -0.6 d^{-2/3}$ 
holds for {\em all }  values of $d$ (in fact, by the above argument, the same
upper estimate is valid for $(\mathbb{E} \|G\|_{op}^r)^{1/r}$ with, say,  
$r = 2$ or  $r=d^{2/3}$).

Going back to the issue of having the precise constant 2 in inequality
(\ref{vrDelta}),  let us note that the other source of difficulty, 
namely the fact that the parameter  $\gamma_k$ is only asymptotically of 
order $\sqrt{k}$ but not equal to $\sqrt{k}$, introduces an error 
that is of smaller order than our ``margin of safety." As pointed out 
earlier, we have $\gamma_k > \sqrt{k-1}$ and so  
$\gamma_k / \sqrt{k} > \sqrt{1 -1/k)} \approx 1 - 1/2k$. The relevant value of
$k$ is $d^2$, leading to the relative error of order $d^{-2}/2$, as
opposed to the margin of safety of $0.3 d^{-2/3}$ (note also that 4 is the
smallest  value of $d$ that is of interest).

\medskip \noindent {\bf Appendix } 
{\bf G } {\em Upper estimates on $\vol \mathcal{S}_N$ for small to moderate
$N$.}  We now indicate how one may use the explicit efficient nets of the sphere
$S^2$ listed in
\cite{HSS} to majorize the volume radius of   $\mathcal{S}_N$ if $N$ is not too
large. As a demonstration, we will derive bounds for the volume of the set of
separable states on 8 qubits (one may say, a qubyte).

The site  \cite{HSS} lists, for $m\in \{4,\ldots,130\}$, sets $\mathcal{N}_m$ of
$m$ points in $S^2$ such that every point of $S^2$ is within $\ep = \ep _m$
(measured in degrees) from one of the points of $\mathcal{N}_m$, with the
dependence $m \raw \ep _m$ ``putatively optimal" (and very likely nearly
optimal). Noting that $\delta = 2\sin \ep/2$
we verify numerically that in most of the interesting range the putatively
optimal value $\delta_m$ verifies $m \, \delta_m^2 \approx 5$ (more precisely,
$5.1 \pm 1\%$, still not far from the asymptotic value $(2/\sqrt{3})^3 \pi
\approx 4.8368$ that we mentioned earlier). Since in the present context the
bound in (\ref{2NlogN}) becomes
$(1-\delta_m^2/2)^{-8} \sqrt{2 \log (2m^{8})}=(\cos \ep_m)^{-8} \sqrt{2 \log
(2m^{8})}$, substituting $m=5/\delta_m^2$ leads to a function $\phi(\delta) 
= (1-\delta^2/2)^{-8} \sqrt{2 \log (2(5/\delta^2)^{8})}$, which attains its
minimum very near $\delta=.15$, which suggests that the optimal value of $m$
should be around 222.  This is beyond the range of the tables from
\cite{HSS}, but using the largest available $m=130$ and the corresponding 
$\ep_{130}=11.3165625^\circ$ yields an upper bound of $10.417406$,
which is less than $2\%$ larger that the majorant that would presumably be
given by
$m=222$. Plugging in the obtained bound into (\ref{mstarpi}) and using the
fact that 
$(\vol \tilde{\Sigma}/{ \vol \Sigma})^{1/d^2} = d^\alpha = (27/16)^{N/8}$ we are
led to 
$$
\upsilon :=({\vol \Sigma}/{ \vol B_{HS} })^{1/d^2} \le 
(16/27) \cdot 10.417406/\gamma_{d^2} < 0.02411446
$$
Taking into account (\ref{SSigma}) and substituting the explicit expression for
$\vol B_{HS}$  we obtain
$$
\vol \cs \le \frac{\sqrt{d}}2  
\frac{\pi^{d^2/2} \  \upsilon^{d^2}}{\Gamma(d^2/2+1)} .
$$
Finally, using the closed formula for $\vol \mathcal{D}$ from (\ref{closed})
we get
$$
\frac{\vol \cs}{\vol \mathcal{D}} \le 
\frac{(2\pi)^{d/2} \  \upsilon^{d^2}\Gamma(d^2)}{2^{d^2/2 \; +1} \
\Gamma(d^2/2+1)\prod_{j=1}^d \Gamma(j)}
< 8.6 \cdot 10^{-19996} ,
$$
which (modulo rounding errors) is equivalent to a much less impressive bound 
$({\vol \cs}/{\vol \mathcal{D}})^{1/\dim \cd}\le 0.49534$. 

The smallest $N$ for which an argument such as above gives a non-trivial upper
bound appears to be 6;  we get $({\vol \cs_6}/{\vol \mathcal{D}_6})^{1/\dim
\cd_6} < 0.95$.  We refer to \cite{slater} and its references for extensive
(largely numerical) treatment of the case $N=2$.

\medskip \noindent {\bf Appendix } 
{\bf H } {\em The in-radii of $\mathcal{S}_N$ and $\Sigma_N$.}
 The papers
\cite{BCJLPS, GB} estimate from {\em below } the in-radius of 
$\mathcal{S}$ in the Hilbert-Schmidt metric by a quantity that is of
order of $d^{-\eta}$, where $\eta = \log{20}/\log{4} \approx 2.160964$ and $3/2$
respectively. The ``trivial" upper bound on that radius is the in-radius
of $\mathcal{D}$, which -- by a rather elementary and well-known argument -- 
equals $1/\sqrt{d(d-1)} = O(d^{-1})$.
By comparing volumes we see that the second inequality in  (\ref{vrSigma})
yields an asymptotically better upper estimate that (up to logarithmic factors)
corresponds to $\eta = 1+ \alpha \approx 1.094361$.   
This follows by taking into account  (\ref{SSigma})
or by observing that,  by simple geometric considerations, the Euclidean
in-radius of $\Sigma$ is at least as large as that of
$\mathcal{S}$ (the latter considered  in the hyperplane 
$H_1$ of trace one matrices).  By  tinkering with the argument it is possible
to remove the logarithmic factors and, indeed, to obtain an upper bound 
on the in-radius of $\Sigma$ (and hence of $\mathcal{S}$) which is 
$o(d^{-1-\alpha})$, but to improve the exponent new ideas would be necessary
 -- if that is at all possible, that is.

Let us also note that our argument  
yields as well a lower bound $6^{-N/2}$ on the in-radius of $\Sigma$ which
corresponds to $\eta = \log{6}/\log{4} \approx 1.3863$, and so is stronger
that those that can be formally derived from \cite{BCJLPS} or \cite{GB}. 
To see this it is enough to combine the ``trivial" lower estimate $d^{-1}=
2^{-N}$ on the in-radius of the set 
$\tilde{\Sigma}_N$  defined in what follows 
(cf. the paragraph following (\ref{vrSigmamod})) with the known value
$(3/2)^{N/2}$  of the norm of the related map $A^{\otimes N}$. However, since
the in-radius of $\mathcal{S}$  may {\em a priori } be (and probably is, except
when $N=1$) smaller than that of $\Sigma$, this  doesn't   
improve the lower bounds from \cite{BCJLPS, GB}. It is also likely
that the methods from those papers may yield a lower bound 
on the in-radius of $\Sigma$ that is better than $6^{-N/2}$.

While our calculations narrow down the potential range of the in-radii of
$\mathcal{S}$ and $\Sigma$, and while further progress along the same lines
may be possible,  it seems
likely that  to obtain the exact values of exponents a more careful calculation
involving, e.g.,  spherical harmonics may be necessary. On the other hand, as
we have already noted, it may be more natural to consider
in this context the inner product norm, which is different from the
Hilbert-Schmidt norm and induced by the inner product $(u, v) \raw (3\, \tr uv
- \tr u \,\tr v)/2$ on each factor $\mathcal{M}_2^{sa}$.   The (solid)
ellipsoid $\cale$, which is the unit ball \wrt the corresponding inner product
norm on 
$\mathcal{M}_d^{sa} = (\mathcal{M}_2^{sa})^{\otimes N}$, verifies $\cale /d
\subset \Sigma
\subset \cale$ (this is just a restatement of $B_{HS}/d \subset
\tilde{\Sigma} \subset B_{HS}$) and these inclusions are essentially optimal.
On the one hand, every pure state clearly belongs to the boundary of $\cale$.
On the other hand, it follows from, say, (\ref{vrSigmamod}) that the volume
radius of 
$\Sigma$ is, on the power scale, the same than that of $\cale /d$.
This means that -- from the volumetric point of view -- $\Sigma$ and $\cale /d$ 
are nearly non-distinguishable. It would be of interest to determine the
precise in-radius of $\mathcal{S}$ \wrt the inner product norm induced by
$\cale$ or, equivalently, the largest $\ep$ such that $I_d/d + \ep (\cale \cap
H_0)  \subset \mathcal{S}$, where 
$H_0 := \{A \in \mathcal{M}_d^{sa}: \tr A = 0 \}$. [It is conceivable that 
that radius is of order $d^{-1}$.]

\medskip \noindent {\bf Appendix } 
{\bf I } {\em Separable states on $N$ qudits.} (See also \cite{qudits}.) Most of
the elements of our analysis can be generalized to tensor products 
involving spaces  ${\mathbb {C}}^D$ with $D > 2$, leading to non-trivial
but not definitive results. As a demonstration, let $D \ge 3$ and consider 
the family of spaces $\calh_N = ({\mathbb {C}}^D)^{\otimes N}$. We shall employ 
analogous notation to that of the main text, in particular 
$d= \dim \calh_N= D^N$. The set of pure states on $\calb({\mathbb {C}}^D)$ 
coincides with the projective space ${\mathbb{C}P}^{D-1}$ 
whose real dimension is $2D-2$ and which admits, for $\delta > 0$, 
$\delta$-nets of cardinality not exceeding $(C'/\delta)^{2D-2}$, where 
$C'$ is a universal constant.  This leads to a bound on $({\vol \Sigma_N}/{ \vol
B_{HS} })^{1/d^2}$ which is of order $(1-\delta)^{-N}\sqrt{ND
\log{(C'/\delta)}}/d$. Choosing, say, $\delta = 1/N$ and using the same bound on
$\vol \mathcal{D}_N$ as earlier combined with the ``easy" part of (\ref{SSigma})
we obtain
$$
({\vol \mathcal{S}_N}/{ \vol \mathcal{D}_N })^{1/\dim \mathcal{D}}
= O\left({\sqrt{ND \log{N}}}/{d^{1/2}}\right) .
$$
Since $d= D^N$, this leads to a non-trivial bound even for $N=2$ if $D$ is large
enough.  It is also possible to improve slightly the exponent of $d$ by 
working (as we did for $D=2$) with a more balanced affine image of $\Sigma_N$.
The resulting improvement $\alpha = \alpha_D$ decreases as  $D$ increases; 
for example, $\alpha_3 = \frac{8 \log{2}}{9 \log{3}}-\frac12 
\approx 0.0608264$ and,  for large $D$,
$\alpha_D \sim (2D \log{D})^{-1}$.  However,  showing optimality of
the so obtained exponents requires -- for $D>2$ -- new ideas and will be
presented elsewhere.

\noindent {\scriptsize Equipe d'Analyse Fonctionnelle, B.C. 186,
      Universit\'{e} Paris VI, 4, Place Jussieu,  F-75252  Paris, France \\
  and \\ 
					Department of Mathematics,
     Case Western Reserve University,
     Cleveland, OH 44106, U.S.A. }\\
    {\footnotesize szarek@ccr.jussieu.fr, szarek@cwru.edu}


\small
\begin{thebibliography}{~~}


{\small
\bibitem{ball} 
K. Ball,
{\em Convex geometry and functional analysis.} 
Handbook of the geometry of Banach spaces, Vol. 1, 
161--194, North-Holland, Amsterdam, 2001. 

\bibitem{BM} J. Bourgain and V. Milman,
{\em New volume ratio  
properties for convex symmetric bodies in $\mathbb{R}^n$.} 
Invent. Math.  88, 319-340.

\bibitem{BC} 
S. L. Braunstein and C. M. Caves, 
{\em  Statistical distance and the geometry of quantum states.}
Phys. Rev. Lett. 72 (1994), no. 22, 3439--3443. 

\bibitem{BCJLPS} 
S. L. Braunstein, C. M. Caves, R. Jozsa, N. Linden, S. Popescu and R. Schack,
{\em   Separability of very noisy mixed states and 
implications for NMR quantum computing}, 
 Phys. Rev. Lett. 83 (1999), 1054-1057

\bibitem{DS}  K. R. Davidson and S. J. Szarek,
{\em Local operator theory, random matrices and Banach spaces.  }
In {\em Handbook of the geometry of Banach spaces,} Vol. 1, 317--366, 
North-Holland, Amsterdam, 2001.
{\em Addenda and Corrigenda}, Vol. 2, 2003, 1819--1820.

\bibitem{GB}   L. Gurvits and H. Barnum, 
{\em Separable balls around the maximally mixed multipartite quantum states. }
http://arxiv.org/abs/quant-ph/0302102

\bibitem{HT}  U. Haagerup and S. Thorbj{\o}rnsen,
{\em Random matrices with complex Gaussian entries. } 
SDU preprint Nr. 7, 1998.

\bibitem{HSS} 
R. H. Hardin, N. J. A. Sloane and W. D. Smith
{\em Spherical Coverings. } 
http://www.research.att.com/\~{}njas/coverings/index.html

\bibitem{HHH}
M. Horodecki, P. Horodecki and R. Horodecki
{\em Mixed states entanglement and quantum communication.}
In {\em Quantum Information: An Introduction to Basic Theoretical 
Concepts and Experiments,} Springer Tracts in Modern Physics, Springer, 
Berlin, 2001.

\bibitem{JL} R. Jozsa and N. Linden,
{\em On the role of entanglement in quantum computational speed-up. }
http://arxiv.org/abs/quant-ph/0201143

\bibitem{ledoux}  M. Ledoux, 
{\em A remark on hypercontractivity and tail 
inequalities for the largest eigenvalues of random matrices. }
S\'eminaire de Probabilit\'es XXXVII. Lecture Notes in Math., 
Springer Verlag, 2003. To appear. Preprint
http://www.lsp.ups-tlse.fr/Ledoux/matrix.ps

\bibitem{L}  M. Ledoux, 
{\em  Isoperimetry and Gaussian analysis. Lectures on probability theory
and statistics (Saint-Flour, 1994)}, 165--294, Lecture Notes in Math.,
1648, Springer, Berlin, 1996.

\bibitem{pisier} G. Pisier,  
{\em The volume of convex bodies and Banach space geometry. }
Cambridge Tracts in Mathematics,  94. 
Cambridge University Press, Cambridge, 1989.

\bibitem{pr} A. O. Pittenger and M. H. Rubin,
{\em Complete separability
and Fourier representations of $n$-qubit states. }
Phys. Rev. A (3) 62 (2000), no. 4, 042306, 5 pp.

\bibitem{prLAA} A. O. Pittenger and M. H. Rubin,
{\em Convexity and the separability problem of quantum 
mechanical density matrices.}
Linear Algebra Appl. 346 (2002), 47--71.

\bibitem{RS} C. A. Rogers and G. C. Shephard,  
{\em The difference body of a convex body. }
Arch. Math. 8 (1957), 220--233.

\bibitem{RS2} C. A. Rogers and G. C. Shephard,  
{\em Convex bodies associated with a given convex body. } 
J. London Math. Soc. 33 (1958), 270--281. 

\bibitem{SR} J. Saint-Raymond,  
{\em Le volume des id\'eaux d'op\'erateurs classiques. }
(French. English summary) [The volume of classical operator ideals] 
Studia Math. 80 (1984), no. 1, 63--75.

\bibitem{qudits} 
P. Rungta, W. J. Munro, K. Nemoto, P. Deuar, G. J. Milburn and C. M. Caves, 
{\em Qudit Entanglement.} 
In {\em Dan Walls Memorial Volume.} Springer, Berlin, 2000.

\bibitem{santalo} L. A. Santal\'o,  
{\em An affine invariant for convex bodies of
$n$-dimensional space. }
(Spanish) Portugaliae Math. 8, (1949), 155--161.

\bibitem{slater} P. B. Slater,
{\em The silver mean and volumes of the separable two-qubit states.}
http://arXiv.org/abs/quant-ph/0308037

\bibitem{SZ} H.-J. Sommers and K. \.Zyczkowski,
{\it Bures volume of the set of mixed quantum states.} 
J. Phys. A: Math. Gen. 36 (2003), 10083--10100.

\bibitem{ST}  
S. J. Szarek and N. Tomczak-Jaegermann,  
{\em On nearly Euclidean
decomposition for some classes of Banach spaces. }
Compositio Math. 40 (1980), no. 3, 367--385.

\bibitem{TW1}  C. A. Tracy and H. Widom, 
{\it Level-spacing distributions and the Airy kernel.}
Comm.\ Math.\ Phys.\  159 (1994), 151--174.

\bibitem{TW2} C. A. Tracy and H. Widom, 
{\it On orthogonal and symplectic matrix ensembles.}
Comm.\ Math.\ Phys.\  177 (1996), 727--754.

\bibitem{VT} G. Vidal and R. Tarrach {\it Robustness of entanglement.}
Phys. Rev. A 59 (1999), 141--155.

\bibitem{ZHSL}  
K. \.Zyczkowski, P. Horodecki, A. Sanpera and M. Lewenstein,
{\it On the volume of the set of mixed entangled states.}
Phys. Rev. A 58 (1998) 883--892.

\bibitem{ZS} K. \.Zyczkowski and H.-J. Sommers, 
{\it Hilbert-Schmidt volume of the set of mixed quantum states.} 
J. Phys. A: Math. Gen. 36 (2003), 10115--10130.

}

\end{thebibliography}
\end{document}